\documentclass[aps,prl,twocolumn,superscriptaddress,showpacs]{revtex4-2}

\usepackage{amsmath,amssymb}
\usepackage{bm}
\usepackage{graphicx}
\usepackage{booktabs}
\usepackage[colorlinks,linkcolor=blue,citecolor=blue,urlcolor=blue]{hyperref}

\newcommand{\bx}{\bm{x}}
\newcommand{\bv}{\bm{v}}
\newcommand{\bs}{\bm{s}}
\newcommand{\btheta}{\bm{\theta}}

\begin{document}

\title{Quantum Dynamics via Score Matching on Bohmian Trajectories}

\author{Lei Wang}
\email{wanglei@iphy.ac.cn}
\affiliation{Beijing National Laboratory for Condensed Matter Physics and Institute of Physics, 
Chinese Academy of Sciences, Beijing 100190, China}

\date{\today}

\begin{abstract}
We solve the time-dependent Schr\"odinger equation by learning
the score function, the gradient of the log-probability density,
on Bohmian trajectories. In Bohm's formulation of quantum
mechanics, particles follow deterministic paths under the
classical potential supplemented by a quantum potential depending on the score function of the evolving density.  These non-crossing Bohmian 
trajectories form a continuous normalizing
flow governed by the score. We parametrize the score with a neural network and
minimize a self-consistent Fisher
divergence between the network and the score of the
resulting density. We prove that the zero-loss minimizer
of this self-consistent objective recovers
Schr\"odinger dynamics for nodeless wave functions, a condition
naturally met in quantum vibrations of atoms. We demonstrate
the approach on wavepacket splitting in a double-well potential
and anharmonic vibrations of a Morse chain.  By recasting real-time quantum
dynamics as a self-consistent score-driven normalizing flow, this
framework opens the time-dependent Schr\"odinger equation to the
rapidly advancing toolkit of modern generative modeling.
\end{abstract}

\maketitle

\paragraph{Introduction---}

The time-dependent Schr\"odinger equation (TDSE) for quantum particles
of mass $m$ with configuration
$\bx \in \mathbb{R}^d$ and potential $V(\bx)$,
\begin{equation}
  i\hbar\frac{\partial\Psi(\bx, t)}{\partial t}
  = \left[-\frac{\hbar^2}{2m}\nabla^2 + V(\bx)\right]\Psi(\bx, t),
  \label{eq:tdse}
\end{equation}
governs quantum dynamics across electronic, atomic, and molecular physics. 
Grid-based methods scale exponentially with the number of degrees
of freedom $d$, motivating alternatives that either expand the wave
function in a time-adaptive basis~\cite{Meyer1990mctdh},
parametrize it with neural networks~\cite{Carleo2017}, or represent
it through particle
trajectories~\cite{LopreoreWyatt1999,SalesMayor1999,Wyatt2005}---the
approach we build on here.

The trajectory framework originates from the Madelung
decomposition~\cite{Madelung1927}, which writes the wave function
as $\Psi = \sqrt{\rho}\,e^{iS/\hbar}$ and recasts
Eq.~\eqref{eq:tdse} as two coupled Eulerian field equations: a
continuity equation for the probability density $\rho$ and a
quantum Hamilton--Jacobi equation for the phase $S$,
\begin{align}
  &\frac{\partial\rho}{\partial t}
    + \nabla\!\cdot\!\bigl(\rho\,\tfrac{\nabla S}{m}\bigr) = 0,
  \label{eq:continuity}\\
  &\frac{\partial S}{\partial t}
    + \frac{|\nabla S|^2}{2m} + V + Q = 0,
  \label{eq:qhj}
\end{align}
where the quantum potential
$Q = -(\hbar^2/2m)\,\nabla^2\!\sqrt{\rho}/\sqrt{\rho}$
encodes all non-classical effects.
De Broglie~\cite{deBroglie1927} introduced the pilot-wave picture
in which a particle is guided by the wave with velocity
$\bv = \nabla S/m$, and Bohm~\cite{Bohm1952a,Bohm1952b} later
independently developed an alternative \emph{interpretation} of quantum
mechanics in terms of hidden variables: the quantum potential $Q$
guides an ensemble of particles along well-defined trajectories
that reproduce all predictions of the Schr\"odinger equation.

In this Letter, we turn Bohmian mechanics from an
interpretive framework into a practical
computational tool for quantum dynamics by revealing its connection to
modern machinery in generative modeling including continuous normalizing
flows~\cite{Chen2018,ZhangEWang2018} and score
matching~\cite{SongErmon2019,Song2021}.

To begin, we define the score function $\bs \equiv \nabla\ln\rho$, in terms
of which the quantum potential takes the form:
\begin{equation}
  Q = -\frac{\hbar^2}{4m}\!\left[
    \nabla\!\cdot\! \bs + \tfrac{1}{2}|\bs|^2
  \right].
  \label{eq:Qscore}
\end{equation}
Writing
$\ln\Psi = \tfrac{1}{2}\ln\rho + iS/\hbar$, one sees that $\bs$ and
$\bv$ are (up to constants) the gradients of the real and imaginary
parts of $\ln\Psi$.  Both are therefore gradient fields by
construction.

\begin{figure}[!tb]
  \centering
  \includegraphics[width=\columnwidth]{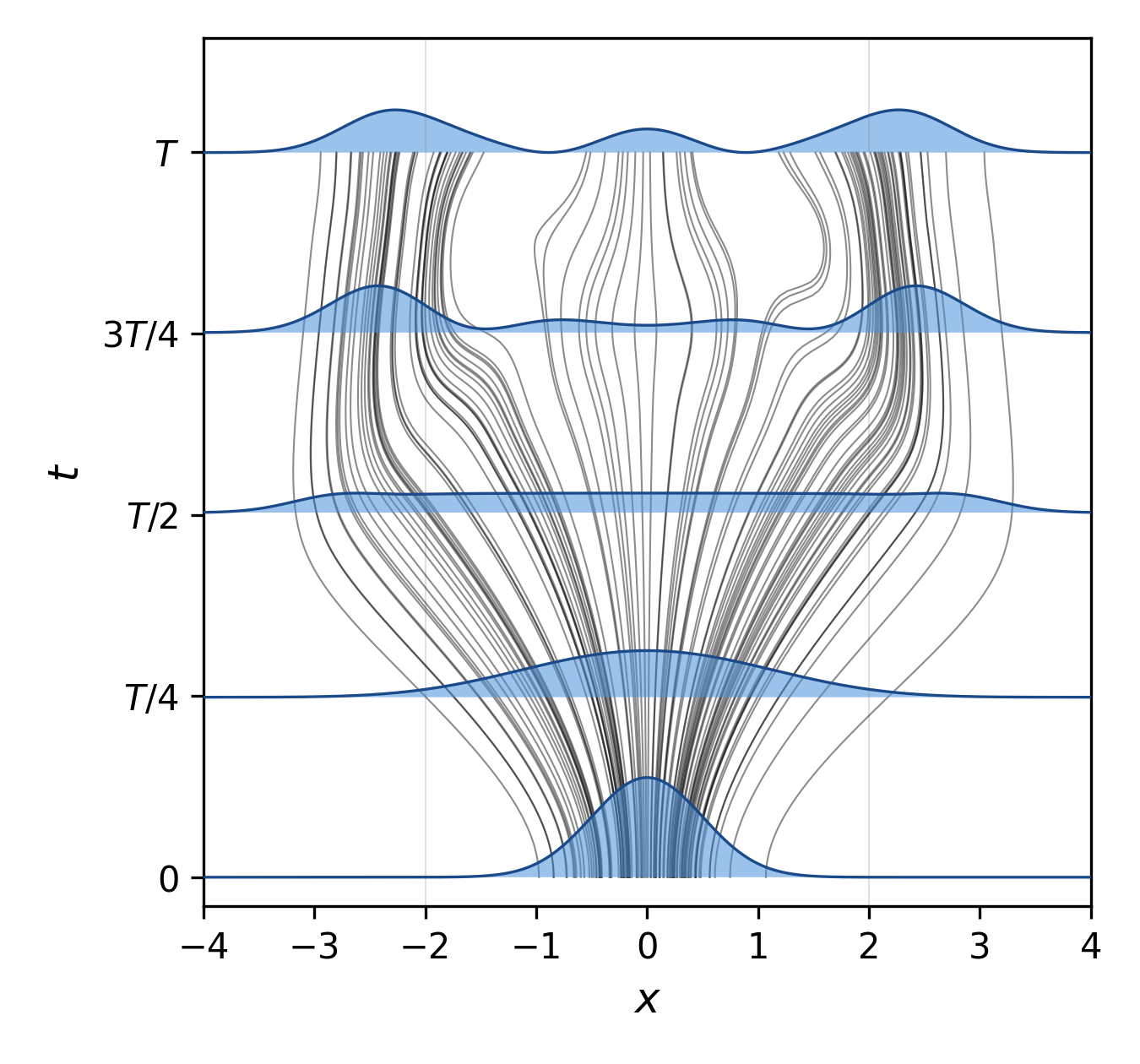}
  \caption{Quantum dynamics as a continuous normalizing flow.
    The black curves are deterministic, non-crossing Bohmian
    trajectories, launched from particle positions sampled from the
    initial density and tracking the time-dependent densities shown
    by the blue waterfall. The blue waterfall plots the exact
    density $|\Psi(x,t)|^2$ at evenly spaced snapshots, showing an
    initial Gaussian wavepacket splitting into multimodal densities
    in a symmetric double-well potential.}
  \label{fig:doublewell}
\end{figure}

To follow individual particles in the Lagrangian frame, we apply
the material derivative
$d/dt = \partial/\partial t + \bv\!\cdot\!\nabla$ to the Madelung
system, obtaining the Bohmian trajectory
equations~\cite{Bohm1952a,Bohm1952b}:
\begin{equation}  
  \frac{d\bx}{dt} = \bv,
  \quad
  m\frac{d\bv}{dt} = -\nabla(V + Q).
  \label{eq:eom}
\end{equation}
This is formally identical to classical mechanics with effective
potential $V + Q$, except that $\bv = \nabla S/m$ is a single-valued
field on $(\bx, t)$ rather than an independent degree of freedom.
By the uniqueness theorem, trajectories cannot cross or pass through
nodes~\cite[{\S}3.3.8]{Holland1993}.  The flow map
$\bx(0) \mapsto \bx(t)$ is therefore a diffeomorphism for nodeless
wave functions, a continuous normalizing flow~\cite{Chen2018},
with $\det F \neq 0$, where
$F(t) = \partial\bx(t)/\partial\bx(0)$ is the deformation gradient
(flow Jacobian).  Since $\bv = \nabla S/m$ is a gradient field, the
flow is more specifically a Monge--Amp\`ere
flow~\cite{ZhangEWang2018}.  Each particle, initialized at
$\bx(0) \sim \rho_0 \equiv \rho(\cdot, 0)$ with velocity
$\bv(0) = \nabla S(\bx, 0)/m$, follows a deterministic trajectory, and the
density evolves exactly as
$\rho(\bx(t),t) = \rho_0(\bx(0))/|\!\det F(t)|$.
Figure~\ref{fig:doublewell} illustrates this flow for a wavepacket
splitting in a one-dimensional double-well: the unimodal initial
density (the base distribution of the normalizing flow) is deformed
into multimodal densities, and non-crossing Bohmian trajectories
launched from random samples of the initial density trace the full
density evolution in between.

In Eq.~\eqref{eq:eom} the phase $S$ no longer appears as a
dynamical field: $\bv$ is propagated directly and $Q$ depends only
on the score $\bs$ via Eq.~\eqref{eq:Qscore}, so the complex phase
need not be represented along the trajectory~\cite{Garashchuk2003,Garashchuk2004}.
The phase enters only through the initial condition $\bv(0) = \nabla S(\bx, 0)/m$
(trivially zero when the initial wave function is real), and $S$ can be reconstructed post
hoc by integrating Eq.~\eqref{eq:qhj} along trajectories if desired.  The practical
difficulty, however, is that integrating Eq.~\eqref{eq:eom} requires
the quantum force $-\nabla Q$, which depends on the score $\bs$ of
the evolving density.  This score is itself unknown and must be
determined self-consistently.
References~\cite{LopreoreWyatt1999,SalesMayor1999} pioneered quantum trajectory methods
by representing the density on a moving mesh and propagating the
hydrodynamic equations directly.
References~\cite{Garashchuk2003,Garashchuk2004}
eliminated the mesh by fitting $\bs$ from the particle ensemble
via a polynomial basis and a functional equivalent to Hyv\"arinen
score matching~\cite{Hyvarinen2005}.  Ref.~\cite{MaddoxBittner2003}
used a Bayesian Gaussian mixture for the same purpose.  These methods fit the score pointwise at each timestep using a
limited basis (polynomials or Gaussian mixtures), without a global
variational principle connecting successive times.

In this Letter, we instead parametrize the score with a neural
network $\bs_{\btheta}(\bx,t)$ and train it end-to-end by minimizing
a single Fisher-divergence objective, simultaneously lifting the
basis-set restriction and enforcing spacetime self-consistency.  The
resulting quantum potential drives a continuous normalizing flow
from the initial density, and any zero-loss minimizer of the
objective recovers the Schr\"odinger solution.

\paragraph{Theory---}

The central idea is to
train the score network by minimizing the mismatch between
$\bs_{\btheta}$ and the score of its own implied density.  The training
objective is the Fisher divergence:
\begin{equation}
  L[\bs_{\btheta}] = \int_0^T\!\! dt
  \int_{\mathbb{R}^d}\! d\bx\,\rho_{\btheta}\,
  \left|\bs_{\btheta} - \nabla\ln\rho_{\btheta}\right|^2,
  \label{eq:fisher}
\end{equation}
where $T$ is the time horizon over which the wave function is
propagated.  Both integrals are estimated on the trajectory itself:
the spatial integral becomes the per-time expectation
$\mathbb{E}_{\bx \sim \rho_{\btheta}(\cdot,t)}\!
\left[|\bs_{\btheta} - \nabla\ln\rho_{\btheta}|^2\right]$
on the particle ensemble, and the time integral becomes a sum over
trajectory checkpoints.  Crucially,
$\rho_{\btheta}$ itself depends on $\bs_{\btheta}$: the score drives
the trajectories that determine the density.  Since $L \geq 0$ with
equality iff $\bs_{\btheta} = \nabla\ln\rho_{\btheta}$, minimizing
$L$ drives the system toward self-consistency.  Because $L$ is an
expectation over $\rho_{\btheta}$, it concentrates on regions where
the physical density has support: the score need not be accurate
where no particles sample.

\phantomsection\label{thm:exactness}\textbf{Theorem}---Suppose the
score $\bs_{\btheta}$ achieves (a)~$L[\bs_{\btheta}] = 0$ and
(b)~induces a Bohmian flow whose density remains strictly positive
on $[0, T]$.  Then
$\bs_{\btheta}(\bx, t) = \nabla\!\ln|\Psi(\bx, t)|^2$, where $\Psi$
is the unique solution of the TDSE~\eqref{eq:tdse} evolving from
the prescribed initial wave function; in particular, the wave
function (and hence its density and phase) is fully recovered.
The proof is detailed in the End Matter.

Evaluating the loss function Eq.~\eqref{eq:fisher} requires $\nabla\ln\rho_{\btheta}$ along each
trajectory.
Differentiating both sides of
$\ln\rho_{\btheta} = \ln\rho_0 - \ln|\!\det F|$
with respect to $\bx(0)$ and transforming to the current frame yields
\begin{equation}
  \nabla_{\bx(t)}\!\ln\rho_{\btheta}
  = F^{-\mathsf{T}}\!\cdot\!
    \bigl[\nabla_{\bx(0)}\ln\rho_0- \nabla_{\bx(0)}\!\ln|\!\det F|\bigr].
  \label{eq:starget}
\end{equation}
Note that the $d\times d$
deformation gradient $F$ is
needed to obtain the target score  $\nabla\ln\rho_{\btheta}$. Differentiating the equations of motion~\eqref{eq:eom} with respect
to the initial position $\bx(0)$ yields the evolution of the
deformation gradient $F$ and its velocity counterpart
$G = \partial\bv(t)/\partial\bx(0)$:
\begin{equation}
  \frac{dF}{dt} = G, \quad
  m\frac{dG}{dt} = -\mathrm{Hess}(V + Q)\cdot F,
  \label{eq:variational}
\end{equation}
with initial conditions $F(0) = I$ and
$G(0) = \partial\bv(0)/\partial\bx(0)$, which reduces to $G(0) = 0$
whenever the initial wave function is real ($\bv(0) = 0$),
as in both applications below.  Here
$[\mathrm{Hess}(V\!+\!Q)]_{ij} = \partial^2(V\!+\!Q)/\partial x_i\partial x_j$
is the Hessian matrix of second derivatives.  $F$ and $G$ track
how an infinitesimal neighborhood of initial conditions deforms
under the flow, and are co-integrated alongside the particle
trajectories using a symplectic leapfrog scheme.

\paragraph{Implementation---}

\begin{figure}[t]
  \centering
  \includegraphics[width=\columnwidth]{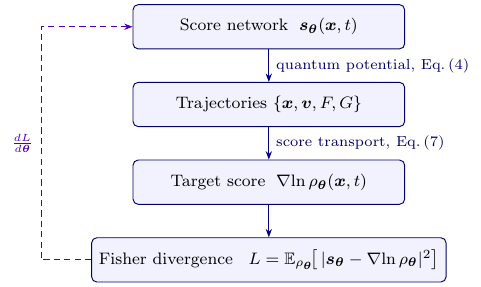}
  \caption{Training loop for the self-consistent score matching on
    Bohmian trajectories.  The score network $\bs_{\btheta}$
    determines the quantum potential $Q$, which together with the
    classical potential $V$ drives the Bohmian trajectories via
    Eq.~\eqref{eq:eom}.  Along each trajectory we co-integrate the
    deformation gradient $F$ and its velocity counterpart $G$ via
    Eq.~\eqref{eq:variational}.  The Fisher divergence
    (Eq.~\eqref{eq:fisher}) measures the mismatch between
    $\bs_{\btheta}$ and the score target (Eq.~\eqref{eq:starget}),
    and gradients $dL/d\btheta$ propagate end-to-end through the
    trajectory via BPTT.}
  \label{fig:framework}
\end{figure}

We parametrize the score function as
$\bs_{\btheta}(\bx,t) = \nabla_{\bx}\varphi_{\btheta}(\bx,t)$, the
gradient of the scalar function $\varphi_{\btheta}$, ensuring $\bs_{\btheta}$ is a gradient field
by construction.  The loss~\eqref{eq:fisher} depends on $\btheta$
through the score $\bs_{\btheta}$, the flow Jacobian $F$, and the
sampling distribution $\rho_{\btheta}$.  Since initial positions
$\bx_i(0) \sim \rho_0$ are fixed, gradients flow end-to-end through
the trajectory ordinary differential equation (ODE) via the chain rule.  The full gradient
$dL/d\btheta$ is computed by backpropagation through time (BPTT)
along the discretized leapfrog trajectory, with gradient
checkpointing~\cite{Chen2016checkpoint} to manage memory.  The
algorithm chains 5 spatial-derivative orders of $\varphi_{\btheta}$,
plus the parameter gradient via BPTT.
The full derivative ladder is given in the End Matter,
Table~\ref{tab:derivatives}.  Gradient stability is aided by the
symplecticity of the leapfrog integrator, which constrains the
growth of Jacobian products.  Figure~\ref{fig:framework} summarizes
the resulting computational loop.

\paragraph{Application---}

We apply the method to a chain of $d$ coupled Morse oscillators
$V = \sum_i D(1 - e^{-\beta x_i})^2 + \lambda\sum_i x_i x_{i+1}$
with $D = 12.5$, $\beta = 0.2$, $\lambda = 0.3$, and the first mode
displaced by $x_0 = 1$.  The score is decomposed as
$\bs_{\btheta} = \bs_\text{base} + \nabla_{\bx}\varphi_{\btheta}$, where
$\bs_\text{base}$ is the analytical score from the harmonic
approximation to the potential, providing a warm start.  The
correction $\varphi_{\btheta}$ is a multi-layer perceptron (MLP)
with Feature-wise Linear Modulation
(FiLM)~\cite{Perez2018film} time conditioning, which modulates
each hidden layer by a learned scale and shift that depend on $t$.  Training
uses $M = 5000$ particles, $T = \pi$, $dt = 0.01$ (314 leapfrog
steps).
Exact Bohmian dynamics on nodeless wave functions admits no
caustics: the quantum potential $Q$ provides the anti-focusing
force that keeps $\det F$ bounded away from zero, and the flow
map is a diffeomorphism (Holland~\cite[{\S}3.3.8]{Holland1993}).
Early in training, however, the learned score yields an
insufficient $Q$ and classical trajectory focusing appears
transiently.  We mask near-caustic timesteps ($|\!\det F| < 0.1$)
in the loss to avoid divergent score targets.  As training
progresses, the learned quantum force $-\nabla Q$ prevents caustic
formation and the masked fraction drops to zero.  This
regularization therefore does not affect the final accuracy.

\begin{figure}[t]
  \centering
  \includegraphics[width=\columnwidth]{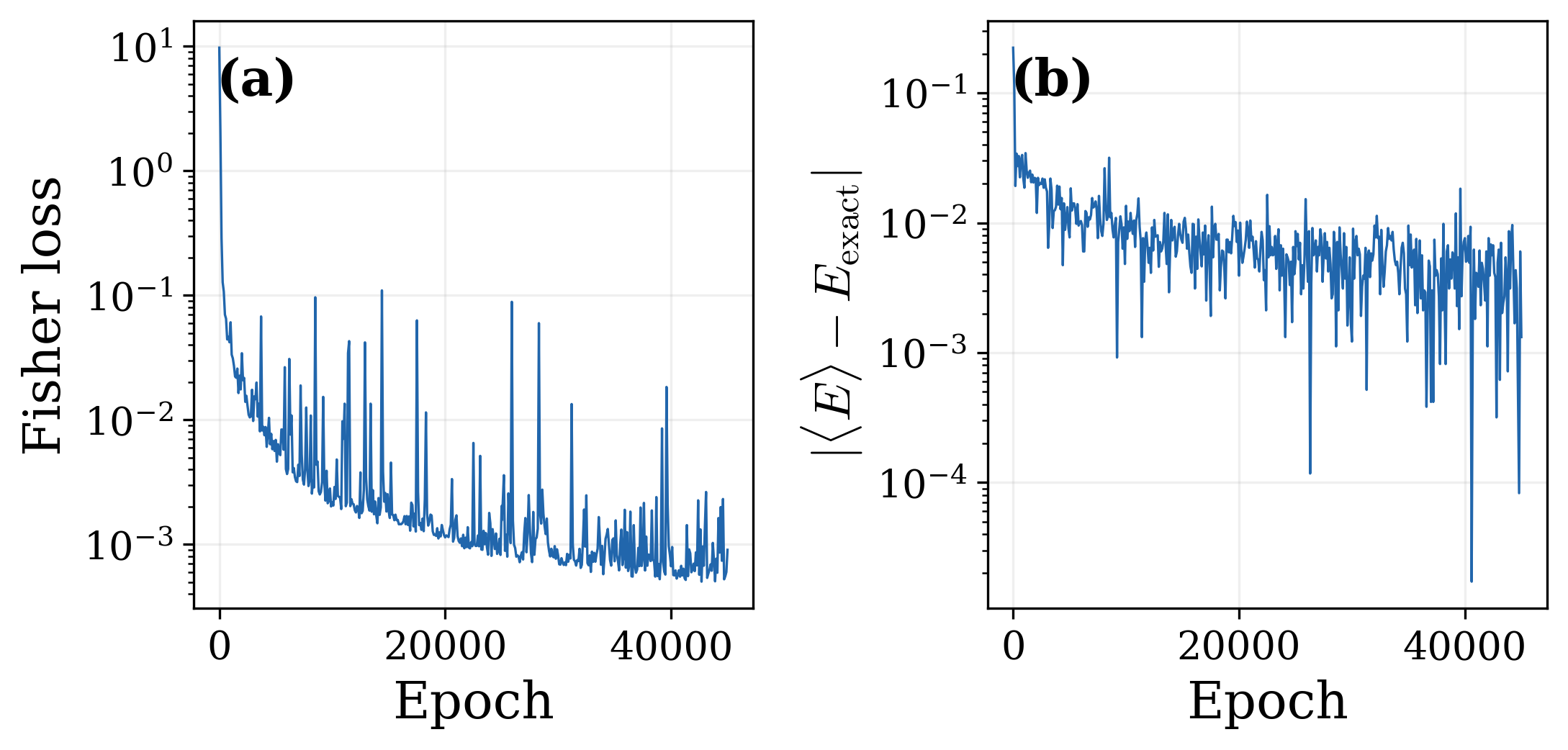}
  \caption{Training convergence on the $d = 4$ Morse chain over
    45{,}000 epochs (raw per-epoch values, log scale).
    (a)~Fisher loss vs.\ epoch.
    (b)~Mean-energy error $|\langle E\rangle - E_\text{exact}|$
    vs.\ epoch.  $\langle E\rangle$ is averaged over the $M = 5000$
    particles and the trajectory time $[0, T]$, and
    $E_\text{exact} = 2.3236$ is obtained from a 4D split-operator
    FFT reference.}
  \label{fig:training}
\end{figure}

Figure~\ref{fig:training} shows the training convergence over
45{,}000 epochs.  The occasional spikes visible in panels~(a)
and~(b) are caustic events: $|\!\det F|$ transiently becomes
small, so the score target
$F^{-\mathsf{T}}[\nabla_{\bx(0)}\ln\rho_0 - \nabla_{\bx(0)}\ln|\!\det F|]$
in Eq.~\eqref{eq:starget} inflates and the network takes an
outsized gradient step.  The system self-heals
within tens of epochs: the learned $Q$ strengthens,
$|\!\det F|$ recovers, and both the loss and energy error return
to baseline.  The Fisher loss [panel~(a)] drops by four
orders of magnitude, from $\sim 10$ at initialization to
$\sim 10^{-3}$.  Panel~(b) reports the mean-energy error
$|\langle E\rangle - E_\text{exact}|$, where the per-particle energy
$\tfrac{1}{2}m|\bv_i|^2 + V(\bx_i) + Q(\bx_i,t)$ is averaged
over all ensemble particles and over the trajectory interval $[0,T]$.
For the exact Bohmian flow, $\langle E\rangle$ at each time equals the
conserved expectation value, so a
single scalar averaged over $[0,T]$ simultaneously probes accuracy
(bias relative to $E_\text{exact}$) and energy conservation (drift
across the interval).
The energy error decreases from $\sim 0.2$ to a median of
$\sim 4\!\times\!10^{-3}$ at convergence, about $0.2\%$ relative to
$E_\text{exact} = 2.3236$ obtained from a 4D split-operator
fast-Fourier-transform (FFT) reference on a $64^4$ grid.

To verify that the trained score reproduces the correct quantum
dynamics, we propagate an independent ensemble of $M = 20{,}000$
particles with the learned score.
Figure~\ref{fig:moments} compares the resulting per-mode mean
$\langle x_i\rangle(t)$ and width
$\sigma_i(t) = \sqrt{\langle x_i^2\rangle - \langle x_i\rangle^2}$
against the exact FFT reference.  Panel~(a) shows dipole-like
oscillations of the mode positions: the initially displaced
$i = 0$ mode oscillates back through and past equilibrium under
its restoring force, while the bilinear coupling
$\lambda\,x_i x_{i+1}$ transmits this excitation along the chain,
driving progressively delayed and smaller-amplitude motion of
modes $i = 1, 2, 3$.  Panel~(b) shows mode-specific breathing of
the widths: the Morse anharmonicity makes the local restoring
force amplitude-dependent, so each marginal is squeezed and
stretched as it samples regions of varying curvature, with
$i = 0$ breathing most strongly because of its larger excursion
and the others inheriting weaker breathing through their coupled
motion.  The network reproduces both effects in phase and
amplitude across all four modes.

\begin{figure}[t]
  \centering
  \includegraphics[width=\columnwidth]{prl_morse_moments.png}
  \caption{Mean $\langle x_i\rangle(t)$ (a) and width
    $\sigma_i(t)$ (b) for each of the four modes of the $d = 4$
    Morse chain.  Solid lines: exact FFT reference.  Open circles:
    $M = 20{,}000$ particles propagated with the trained score
    network.}
  \label{fig:moments}
\end{figure}

\begin{figure}[t]
  \centering
  \includegraphics[width=\columnwidth]{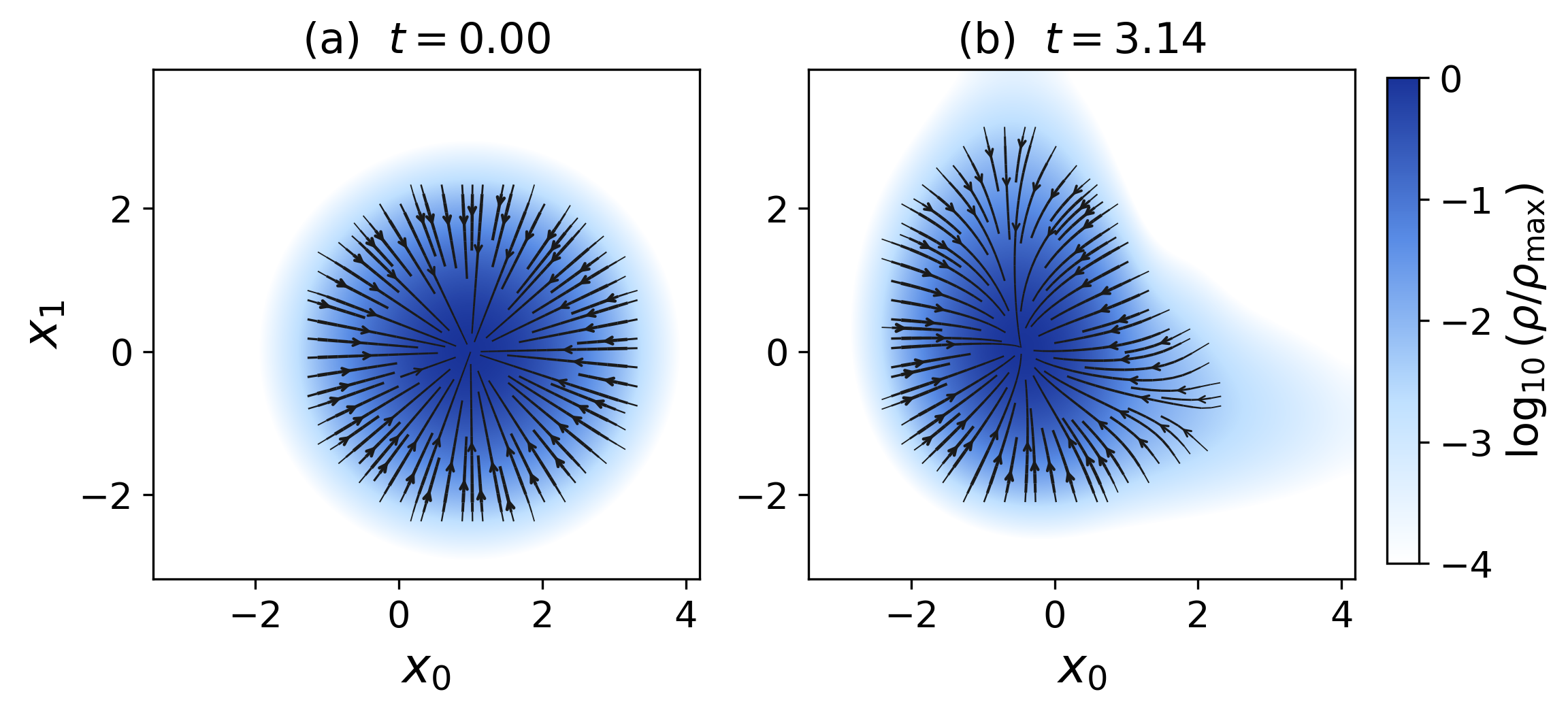}
  \caption{Learned score field $\bs_{\btheta}(\bx, t)$ (streamlines)
    on the $(x_0, x_1)$ slice of the $d = 4$ Morse chain
    ($x_2 = x_3 = 0$).  Background: exact slice density on a log
    scale.
    (a)~$t = 0$: Gaussian initial state with approximately radial
    score.
    (b)~$t = \pi$: anharmonic coupling deforms the density and the
    score field.  The learned streamlines remain aligned with the
    exact density gradient across the wavepacket.}
  \label{fig:score_field}
\end{figure}

To visualize what the network has learned, Fig.~\ref{fig:score_field}
shows the score field $\bs_{\btheta}(\bx, t)$ on the $(x_0, x_1)$
slice through the 4D configuration space, with non-plotted dimensions
fixed at equilibrium ($x_2 = x_3 = 0$).  The background color shows
the exact slice density
$\rho(x_0, x_1, 0, 0) = |\Psi(x_0, x_1, 0, 0)|^2$ on a logarithmic
scale, computed from the FFT wave function.  Streamlines depict the
learned score field $\bs_{\btheta}$, which the network must produce
self-consistently to satisfy $\bs_{\btheta} = \nabla\ln\rho_{\btheta}$.

At $t = 0$ [panel~(a)], the score field is approximately radial,
reflecting the Gaussian initial state.  At $t = \pi$ [panel~(b)], the
density has deformed significantly under the anharmonic coupled
potential: the wavepacket elongates along $x_0$ and loses its
near-rotational symmetry.  The score field tracks
this deformation, curving to follow the density gradient of the
non-Gaussian distribution.  The network captures the full 4D score on
these trajectories: even though only a 2D projection is displayed,
$\bs_{\btheta}$ is a 4-component vector field that must simultaneously be
correct in all dimensions to maintain the self-consistent density
evolution.
As a direct density-level check, we evaluate the reverse
Kullback--Leibler divergence
$\mathrm{KL}(\rho_{\btheta}\|\rho_\text{exact})(t)
 \approx M^{-1}\sum_i[\ln\rho_{\btheta}(\bx_i(t),t) - \ln\rho_\text{exact}(\bx_i(t),t)]$
as a per-particle Monte-Carlo average, using
$\ln\rho_{\btheta}(\bx_i(t),t) = \ln\rho_0(\bx_i(0)) - \ln|\!\det F_i(t)|$
from the change of variables and $\rho_\text{exact}=|\Psi|^2$
multilinearly interpolated from the 4D FFT reference.  At $M = 20{,}000$
the KL remains near the interpolation floor throughout the trajectory
($\sim 10^{-2}$ nats at $t = 0$ and $t = \pi$), confirming that the
learned density matches the Schr\"odinger solution at the full
density level, not only in low-order moments.

\paragraph{Discussion---}

Bohm's theory was famously regarded as a conceptual framework to interpret
quantum mechanics~\cite{Bohm1952a,Bohm1952b}. Self-consistent score matching
on Bohmian trajectories turns it into a practical computational tool for the
TDSE, leveraging machinery from modern generative modeling. The link between
score matching and quantum mechanics in fact predates that machinery itself.
Already in 2003, Garashchuk and Rassolov~\cite{Garashchuk2003,Garashchuk2004}
fitted the quantum potential on trajectories by minimizing
$\mathbb{E}_\rho[\tfrac{1}{2}|\bs|^2 + \nabla\!\cdot\!\bs]
\propto \langle Q\rangle$ in a polynomial basis --- the implicit score-matching
loss of Hyv\"arinen~\cite{Hyvarinen2005} up to constants.  What modern generative modeling
contributes---which we leverage here---is a flexible neural
parametrization of the score, automatic differentiation through long
trajectories, and a globally enforced self-consistent objective over
the entire time horizon, rather than an instantaneous least-squares
fit at each timestep.

The parallel between generative modeling and quantum mechanics
extends from deterministic trajectories to stochastic ones.
Nelson's stochastic mechanics~\cite{Fenyes1952,Nelson1966} is a
Schr\"odinger-equivalent formulation in which particles follow a
drift--diffusion process rather than the deterministic Bohmian
path; the drift combines the Bohmian velocity with an osmotic term
proportional to the score, $\bv + (\hbar/2m)\nabla\ln\rho$.  In
modern generative modeling, flow-based models transport densities
along deterministic ODEs, while diffusion-based
models~\cite{SongErmon2019,Song2021} use stochastic differential
equations (SDEs) with a score-proportional drift, the two linked by
a probability-flow ODE that reproduces the SDE's marginals.  The Bohmian ODE is precisely
the probability-flow ODE of Nelson's SDE: Bohmian mechanics is to
flow-based generative models what Nelson's stochastic mechanics is
to diffusion-based ones.  The same viewpoint underlies recent
neural methods for the Fokker--Planck
equation~\cite{Maoutsa2020,Shen2022,Boffi2023,Li2023scvm,WuLiao2025},
which replace SDE simulation by integration of a deterministic ODE
driven by a self-consistently learned score or velocity field.  We
extend this construction to Schr\"odinger dynamics.
The
broader toolkit of modern generative modeling, including denoising
score matching and flow
matching~\cite{Lipman2023,Liu2023rectified,Albergo2023}, may extend
naturally to quantum dynamics. Orlova~\emph{et al.}~\cite{Orlova2024dsm}
have taken a first step on the stochastic side with Deep Stochastic
Mechanics: they parametrize \emph{both} the current and osmotic
velocities of Nelson's process as neural networks, sample
trajectories self-consistently by Euler--Maruyama integration of the
learned SDE, and train the networks by minimizing partial
differential equation (PDE) residuals of the corresponding velocity
equations on those samples.  The sampling
strategy is thus self-generated as in the present method, but the
governing dynamics are stochastic rather than deterministic, two
velocity fields are learned instead of a single score, and the loss
is a PDE residual rather than the Fisher divergence.

Other neural approaches to the TDSE differ more sharply.
Time-dependent variational Monte Carlo
(tVMC)~\cite{Carleo2012,Carleo2017}, among the most established
neural-variational methods for quantum dynamics, parametrizes the
complex wave function and evolves its parameters in time by
projecting the TDSE onto the tangent space of the parameter
manifold via the time-dependent variational
principle~\cite{Dirac1930tdvp,McLachlan1964tdvp}.  This projection
requires inverting an often ill-conditioned quantum geometric
tensor at every integrator step, together with Monte Carlo sampling
from the time-dependent wave function.  The present method trades
nodal flexibility for a real-valued, density-only representation that
is free of sign and phase, replaces the instantaneous tangent-space
projection by a single global loss over the time horizon, replaces
Metropolis sampling of particle coordinates at every step by a deterministic
particle ensemble transported along the Bohmian trajectories, and
needs no metric inversion at all.

Physics-informed neural networks
(PINNs)~\cite{Raissi2019pinn} are another widely used neural
framework for solving PDEs that has been applied to the TDSE.
PINNs parametrize the complex space--time wave function
$\Psi_{\btheta}(\bx, t)$ directly and minimize the TDSE residual
on a set of collocation points drawn from a user-prescribed
distribution over configuration space and time.  As in the present
method, the optimization is a single global loss over the time horizon,
but the collocation distribution is chosen by hand rather than by
the dynamics: equal weight is given to regions the physical state
never visits, and accuracy degrades in the tails where
$|\Psi|^2$ is small.
A recent application of PINNs to the fermionic
TDSE~\cite{Hou2026fastnet} samples collocation from
$|\Psi_{\btheta}|^2$ at each fixed $t$ to bring the training
distribution onto the physical support. A related global-in-time scheme~\cite{kqvx-dl54} expands
the complex wave function in a fixed neural-quantum-state basis
with time-dependent coefficients, and likewise samples from
$|\Psi_{\btheta}|^2$ at each time-integration node. The present
method drops the complex phase via the Madelung passage and
replaces user-prescribed collocation or $|\Psi_{\btheta}|^2$
sampling by a self-generated particle ensemble transported along
the Bohmian trajectories, automatically allocating the training
effort where the density has support.

Closer in spirit
to the present method is the parametrized Wasserstein Hamiltonian flow (WHF)
of Ref.~\cite{Wu2025whf}, which likewise eliminates the
complex wave function and casts Schr\"odinger dynamics as a continuous
normalizing flow on the probability density.  Like tVMC, however,
WHF projects the TDSE onto the parameter manifold at each instant
and integrates the resulting ODE for the neural-network parameters
step by step. The present method
replaces this integrator-driven projection by a global optimization
of the Fisher divergence over the time horizon, endowed with a
built-in convergence diagnostic ($L = 0$ iff self-consistent).

Fermionic wave functions generically have nodes: antisymmetry under
particle exchange forces $\Psi = 0$ on a set that partitions
configuration space into sign-disconnected
cells~\cite{Ceperley1991nodes}.  These nodes violate the
$\rho_{\btheta} > 0$ hypothesis of the present analysis on two
counts.  First, at a node, the score $\bs = \nabla\ln\rho$ diverges
and Bohmian trajectories cannot cross.  A smooth score network
$\bs_{\btheta} = \nabla\varphi_{\btheta}$ cannot represent the
resulting cell structure.  Second, nodes resurrect the Wallstrom
objection~\cite{Wallstrom1994}: the Madelung system recovers the
Schr\"odinger equation only under the quantization condition
$\oint\bv\cdot d\bm{\ell} = 2\pi n\hbar/m$, $n\in\mathbb{Z}$, on
every closed loop in configuration space, a constraint automatic
for a single-valued $\Psi$ but invisible to a score-only
formulation.  A possible route that resolves both issues is to employ the
fermion flow
architecture~\cite{Xie2022,Xie2023mstar,Li2026hugoniot}, which
parametrizes many-electron wave functions as a
permutation-equivariant normalizing flow from a Slater
determinant. A diffeomorphism maps nodal cells
to nodal cells, so within each cell $\rho > 0$ and the phase-free
dynamics of Eq.~\eqref{eq:eom} applies.  The sign across cells is
read off the Slater determinant rather than learned by the score
network.  The Slater determinant is also single-valued with integer
phase winding fixed by its orbitals, and a diffeomorphism preserves
that winding on every closed loop, so the Wallstrom condition is
automatically satisfied.  Combining such fermionic flow
architectures with self-consistent score matching is a promising
path toward real-time quantum dynamics of many-electron systems.

The code and trained model weights that reproduce the results of this
Letter are publicly available at
\url{https://github.com/wangleiphy/BohmianFlow}.


\begin{acknowledgments}
The author acknowledges discussions with Tao Wang,
Yuan Wan, Kai Zhou, Han Wang, Linfeng Zhang, Zongyue Liu, and Marín Bukov.
This work is supported by the
National Natural Science Foundation of China under Grants No. T2225018, No. 12188101, No. T2121001, and
the Cross-Disciplinary Key Project of Beijing Natural Science Foundation No. Z250005.

\end{acknowledgments}

\bibliography{prl_fisher_bohmian}

\clearpage
\onecolumngrid
\begin{center}
  \textbf{\large End Matter}
\end{center}
\vspace{0.5em}
\twocolumngrid

\appendix

\section{Proof of the exactness theorem}
\label{app:proof}

We prove \hyperref[thm:exactness]{the Exactness Theorem} by
constructing the wave function
$\Psi = \sqrt{\rho_{\btheta}}\,e^{iS/\hbar}$ explicitly from
$\bs_{\btheta}$ and verifying that it satisfies the
TDSE~\eqref{eq:tdse}.  The construction also provides the practical
recipe for recovering the density and phase from a trained score.

(i)~\emph{Self-consistency.}~The network $\bs_{\btheta}$ and the
score $\nabla\ln\rho_{\btheta}$ of the density $\rho_{\btheta}$
that it induces through the flow driven by $Q[\bs_{\btheta}]$ are,
\emph{a priori}, distinct fields.  Since $L \geq 0$ with equality
iff its integrand vanishes on the support of $\rho_{\btheta}$,
hypothesis~(a) forces
$\bs_{\btheta}(\bx,t) = \nabla\ln\rho_{\btheta}(\bx,t)$ wherever
$\rho_{\btheta} > 0$, identifying $Q[\bs_{\btheta}]$ with the
Madelung quantum potential $Q[\nabla\ln\rho_{\btheta}]$.

(ii)~\emph{Density.}~The Bohmian flow driven by $\bs_{\btheta}$
pushes $\rho_0$ forward to
$\rho_{\btheta}(\bx(t),t) = \rho_0(\bx(0))/|\!\det F(t)|$ --- the
recipe for the density.  Probability is therefore exactly conserved
on Lagrangian particles, and the Madelung continuity
equation~\eqref{eq:continuity} holds by construction.

(iii)~\emph{Existence of a phase.}~The initial velocity
$\bv(0) = \nabla S(\bx, 0)/m$ is irrotational.  Expanding the
material derivative, the Bohmian equation
$m\,d\bv/dt = -\nabla(V+Q)$ reads
$\partial_t\bv = -\nabla(V+Q)/m - (\bv\!\cdot\!\nabla)\bv$.  Taking
the curl and using the vorticity identity
$\nabla\!\times\![(\bv\!\cdot\!\nabla)\bv] =
(\bv\!\cdot\!\nabla)\bm{\omega} - (\bm{\omega}\!\cdot\!\nabla)\bv
 + \bm{\omega}(\nabla\!\cdot\!\bv)$
with $\bm{\omega} \equiv \nabla\!\times\!\bv$ yields the standard
vorticity equation
$d\bm{\omega}/dt = (\bm{\omega}\!\cdot\!\nabla)\bv - \bm{\omega}(\nabla\!\cdot\!\bv)$,
whose right-hand side is linear in $\bm{\omega}$.  Hence
$\bm{\omega}(0) = 0$ is preserved along trajectories.  A
single-valued phase $S$ with $\bv = \nabla S/m$ therefore exists
for all $t \in [0, T]$.

(iv)~\emph{Quantum Hamilton--Jacobi equation.}~Substituting
$\bv = \nabla S/m$ into
$\partial_t\bv + (\bv\!\cdot\!\nabla)\bv = -\nabla(V+Q)/m$ and using
the irrotational identity
$(\bv\!\cdot\!\nabla)\bv = \tfrac{1}{2}\nabla|\bv|^2$ (from the
general identity
$(\bv\!\cdot\!\nabla)\bv = \tfrac{1}{2}\nabla|\bv|^2 + (\nabla\!\times\!\bv)\!\times\!\bv$
together with $\nabla\!\times\!\bv = 0$) gives
$\nabla\bigl[\partial_t S + |\nabla S|^2/(2m) + V + Q\bigr] = 0$, so
Eq.~\eqref{eq:qhj} holds up to a gauge function $C(t)$ absorbed by
$S \to S - \int_0^t C(t')\,dt'$.  Numerically, $S$ is reconstructed
along each trajectory by integrating Eq.~\eqref{eq:qhj} from the
initial value $S(\bx, 0)$ --- the recipe for the phase.

By hypothesis~(b), $\rho_{\btheta} > 0$ on $[0, T]$, so $\Psi$ is
nodeless and the phase $S$ just constructed is globally
single-valued.  The Madelung
equations~\cite{Madelung1927} [Eqs.~\eqref{eq:continuity}--\eqref{eq:qhj}]
are both satisfied, hence $\Psi$ solves the TDSE~\eqref{eq:tdse},
and step~(i) gives
$\bs_{\btheta} = \nabla\ln\rho_{\btheta} = \nabla\!\ln|\Psi|^2$.
\hfill$\square$

\section{Derivative chain and dimensional scaling}
\label{app:scaling}

Table~\ref{tab:derivatives} lists the 5 spatial-derivative orders
of the learnable scalar potential $\varphi_{\btheta}$ that appear in
the forward pass, plus the parameter gradient via BPTT that yields
the loss gradient $dL/d\btheta$.  Each row of the loss cascade adds
one stage: score, quantum potential, quantum force, Hessian of
$V+Q$ driving the variational equations, the score-transport
contribution $\nabla_{\bx(0)}\ln|\!\det F|$, and finally BPTT
through the full trajectory.

\begin{table}[h]
\caption{Derivative chain of the learnable scalar potential
  $\varphi_{\btheta}(\bx,t)$.  Each stage adds one order of
  differentiation.}
\label{tab:derivatives}
\begin{tabular}{@{}ccl@{}}
  \toprule
  Derivative Order & Quantity & Comment \\
  \midrule
  1st & $\bs = \nabla\varphi$ & Score function \\
  2nd & $\nabla\!\cdot\!\bs$ & Quantum potential, Eq.~\eqref{eq:Qscore} \\
  3rd & $-\nabla Q$ & Quantum force, Eq.~\eqref{eq:eom} \\
  4th & $\mathrm{Hess}(V\!+\!Q)$ & Eq.~\eqref{eq:variational} \\
  5th & $\nabla_{\bx(0)}\!\ln|\!\det F|$ & Score transport, Eq.~\eqref{eq:starget} \\
  6th & $dL/d\btheta$ & BPTT \\
  \bottomrule
\end{tabular}
\end{table}

In $d$ spatial dimensions, the per-step cost is $O(d^2)$ for
propagating the deformation gradient $F$ and its velocity
counterpart $G$ via the variational equations.  The score-transport
term $\nabla_{\bx(0)}\ln|\!\det F|$ can be evaluated by $d$
forward-mode tangent passes ($O(d^3)$ total) or a single reverse-mode
pass ($O(d^2)$). 
This $O(d^2)$ overhead is fundamental to second-order dynamics.
In first-order probability flows for the Fokker--Planck
equation~\cite{Maoutsa2020,Shen2022,Boffi2023,Li2023scvm,WuLiao2025},
the score enters the velocity directly through the probability-flow
ODE, the velocity gradient
$\partial\bv/\partial\bx$ is a local Jacobian of a neural network,
and the Hutchinson trace estimator~\cite{Chen2018,Grathwohl2019}
recovers $O(d)$ scaling.  In Bohmian dynamics, the score enters one
derivative deeper, through the acceleration
$\ddot{\bx} = -\nabla(V + Q(\bs))$, so the velocity gradient
$\partial\bv/\partial\bx = G F^{-1}$ is a trajectory-dependent
quantity that requires the full $d\times d$ deformation gradient.

%

\end{document}